# Effect of Stacking Fault Energy on Mechanism of Plastic Deformation in Nanotwinned FCC Metals

Valery Borovikov[1], Mikhail I. Mendelev[1], Alexander H. King[1,2] and Richard LeSar[1,2]

[1]*Division of Materials Sciences and Engineering, Ames Laboratory, Ames, IA 50011*

[2]*Department of Materials Science and Engineering, Iowa State University, Ames, IA 50011*

**Abstract**

Starting from a semi-empirical potential designed for Cu, we developed a series of potentials that provide essentially constant values of all significant (calculated) materials properties except for the intrinsic stacking fault energy, which varies over a range that encompasses the lowest and highest values observed in nature. These potentials were employed in molecular dynamics (MD) simulations to investigate how stacking fault energy affects the mechanical behavior of nanotwinned face-centered cubic (fcc) materials. The results indicate that properties such as yield strength and microstructural stability do not vary systematically with stacking fault energy, but rather fall into two distinct regimes corresponding to "low" and "high" stacking fault energies.

# 1. Introduction

The stacking fault energy (SFE) plays a critical role in the deformation properties of face-centered cubic (fcc) metals and alloys. The SFE influences such important phenomena as the formation of partial dislocations, the ability of a dislocation to cross slip, and the formation of twin boundaries, all of which having an effect on yield behavior [1, 2]. As attention has been increasingly focused on nanocrystalline and nanotwinned materials because they show very high strength along with good ductility, thermal stability and electrical conductivity at room temperature [3, 4], the stacking fault energy has again drawn attention, especially with regard to critical size effects in which the splitting distance between the partials is comparable to the characteristic dimensions of the microstructure. Recent computer simulations, for example, have shown a strong dependence on the critical grain size for "inverse Hall-Petch" behavior, in which the strength of a polycrystalline material shows a transition from hardening to softening with decreasing grain size [5]. It was shown that there is a crossover from dislocation-driven to grain boundary-mediated deformation with decreasing grain size that depends on a combination of the stacking-fault energy, the elastic properties of the material, and the magnitude of the applied stress [6].

Despite its importance, it is difficult to isolate the specific contribution of the stacking fault energy on mechanical behavior because we have few, if any, examples of materials with stacking fault energies that are different, but with all of their other materials properties, such as unstable stacking fault energy or elastic moduli, being essentially the same. The authors of [7] proposed to use the ratio of the unstable and stable stacking fault energies to compare the results for different metals but other relevant materials properties were still different in that case. In the present paper, we show that one can construct interatomic potentials such that the stacking fault energy can be varied while keeping other calculated properties, such as the unstable stacking fault energy, elastic moduli, cohesive energy, and surface energies, essentially constant. We then employ classical molecular dynamics (MD) simulations to explore the effect of varying the stacking fault energy on the deformation behavior of a structure that approximates what is seen in nanotwinned fcc metals.

It is generally a challenge to modify a potential to change a single material property while leaving all other material properties unchanged, because all of a material's structural and mechanical properties are affected by the potential. We found, however, that a potential first developed to simulate the solid-liquid interface properties in pure Cu [8], in which the potential was not explicitly fit to the SFE energy, could be modified to fit the SFE [9] without leading to significant changes in the reproduction of other target properties. We use the same procedure to develop a series of semi-empirical potentials that produce different SFE energies while leaving other important properties essentially unchanged, which enables us to isolate and study the effect of the SFE on the mechanism of plastic deformation in nanotwinned fcc metals. The rest of the paper is organized as follows. First, we present the developed semi-empirical potentials. Next, we describe the MD simulation scheme we used to test the mechanical behavior of nanotwinned fcc materials. Finally, we discuss the results showing the effects of the stacking fault energy on the mechanical behavior.

## 2. Semi-empirical potentials

The most widely used semi-empirical model for describing interatomic interactions in fcc metals is the embedded atom method (EAM) [10]. The total energy in this method is divided into two contributions: a pair interaction and an embedding energy:

$$U = \sum_{i=1}^{N-1}\sum_{j=i+1}^{N} \varphi(r_{ij}) + \sum_{i=1}^{N} \Phi(\rho_i),  \qquad (1)$$

where the subscripts $i$ and $j$ label atoms, $N$ is the number of atoms in the system, $r_{ij}$ is the separation between atoms $i$ and $j$, $\varphi$ is the pair potential and $\Phi$ represents the energy to embed an atom in a background charge density, $\rho_i$:

$$\rho_i = \sum_j \psi(r_{ij}), \qquad (2)$$

where $\psi$ is the contribution to the electron density from the neighboring atom $j$. Thus, three functions, $\varphi(r)$, $\psi(r)$ and $\Phi(\rho)$, are required to define an EAM potential model for a single component system.

In the present study, we used the same function $\psi(r)$ for all potentials and $\varphi(r)$ and $\Phi(\rho)$ were fitted to reproduce the target properties (listed in Reference [8]), which included T=0 crystal properties, melting point data and liquid pair correlation functions. We then created a series of potentials using different target values for the stacking fault energy, keeping all other target values constant. The developed potentials are labeled as MCux, with x ranging from 1 to 7, their names being ordered according to the value of the target SFE. MCu3 is the potential developed in Reference [9]. All the potentials can be found in [11, 12].

The functions of the developed potentials are shown in Fig. 1. Probably the easiest way to understand how these potentials lead to different values of the SFE is to analyze the shape of the effective pair potential defined as

$$\varphi^{eff}(r) = \varphi(r) + 2\left.\frac{\partial \Phi}{\partial \rho}\right|_{\rho_0} \psi(r), \qquad (3)$$

where $\rho_0$ is the value of $\rho$ in equilibrium at zero temperature. Figure 1 shows that to increase the SFE the potential should have a deeper minimum compensated by the positive branch at larger atomic separations.

The properties obtained with the potentials are presented in Table I. Examination of this table shows that the developed potentials indeed provide a wide range of values for the stacking fault energy while leaving most of the properties unchanged. The calculated values for the SFE range between 0.91 meV/Å² for MCu1 to 11.65 meV/Å² for MCu7. For comparison, the experimental values of the stacking fault energies range between 1.00 meV/Å² (Ag) and 10.36 meV/Å² (Al) [1]. There are, of course, some properties that should correlate with the SFE (e.g., coherent twin boundary (CTB) energy or the difference between hcp and fcc energies), which are different for the various potentials. There is also a clear trend in the free surface energy, which decreases with increasing SFE, and a very weak correlation between the elastic constants and SFE (with a maximum variation less than 3% across the 7 potentials).

We can employ other structural properties to shed further light on the different potentials. In Figure 2, we see that the potentials lead to slightly different temperature dependence of the lattice parameter, which indicates some small deviations in the curvature of the potential surface near the equilibrium positions. Figure 3, however, shows that the potentials lead to almost the same liquid pair correlation function at 1973 K, which indicates that the overall potential surfaces are very similar at that temperature. All potentials lead to about the same value of the unstable stacking fault energy: 15.6 meV/Å$^2$ [13]. There are very weak trends in the vacancy formation and migration energies with increasing SFE (see Table I). If we ignore these weak variations we can suppose that the potentials may lead to about the same activation energy for other processes as well. Overall, the only calculated property of relevance to this study with any significant variation is the stacking fault energy.

## 3. Molecular dynamics simulation of plastic deformation

Molecular dynamics simulations of plastic deformation were performed using the LAMMPS parallel simulation code for systems with a large number of atoms [14]. The simulation cell contained two parallel sets of twins, separated by grain boundaries, as shown in Fig. 4. In the images of the simulation cells, provided throughout this paper, the atoms are colored according to the common neighbor analysis (CNA) [15, 16]. The CNA assigns a structure type (fcc, bcc, hcp, etc.) to every atom. The algorithm is based on a nearest-neighbor graph that encodes the bond connectivity among neighbors of a given central atom. Analysis and visualization of MD simulation snapshots were performed using the software package OVITO (Open Visualization Tool) [17].

The simulation cells were constructed by joining two parts of the system, each containing coherent twin boundaries (indicated here by CTB), which had been tilted around the y-axis (<110>) (see Fig. 4). The tilt angle was $\alpha \approx 27.47°$. The simulation cell dimensions are 444 Å × 129 Å × 484 Å in the x, y and z directions, respectively, with the total number of atoms in the system N = 2,285,000. Periodic boundary conditions were applied in all directions. After joining the two parts, the whole system was relaxed at T=0 K with periodic boundary conditions applied in all directions. The separation distance between the CTBs was ~10.5 nm. We note that this microstructure is similar to as-sputtered microstructures produced and tested experimentally [18]. However, the simulation system we use is simplified, compared to fully three-dimensional columnar-grained experimentally produced samples with preferentially oriented nanotwins, in that the structure is highly uniform and the twin and grain boundaries are arranged with specific symmetries relative to the principle tensile stress axis.

In addition to the CTBs, there are three distinct grain boundaries (GBs) present in the simulation system. If we arbitrarily designate the alternating crystal orientations in each grain as "matrix" and "twin" then the GB parameters depend on whether the abutting crystals are "matrix-matrix," "twin-twin", "twin-matrix," or "matrix-twin." The GBs corresponding to "matrix-matrix" and "twin-twin" cases are asymmetric tilt GBs (ATGBs) with misorientations of $2\alpha \approx 54.94°$ and are identical to each other except for a rotation of 180º about the z-axis. The "matrix-twin" and "twin-matrix" cases produce distinct symmetric tilt GBs (STGBs) with misorientations of $|2\alpha + 70.53|° \approx 35.47°$ and $|2\alpha - 70.53|° \approx 15.59°$.

After the initial relaxation at T=0, the models were annealed at T=1200 K for 4 ns using isothermal-isobaric (NPT) ensemble. All systems were then cooled to T=300 K and tensile loading with a constant strain rate of $10^8$ s$^{-1}$ was applied in the x-direction. The stresses in the y and z directions were kept zero. The stress-strain curves calculated with the set of potentials are shown in Fig. 5. This figure suggests that the potentials can be divided into two groups based on the deformation behavior: the potentials of the first group (MCu1-MCu5) lead to much higher peak stresses than do the potentials of the second group (MC6 and MCu7). As we shall see, there are distinct differences in the physical mechanisms that lead to the deformation behavior

## 4. Results and Discussion

Figure 6 shows MD simulation snapshots obtained using the potentials MCu1 and MCu7, which have the smallest and the largest SFEs, respectively 0.91 meV/A$^2$ for MCu1 and 11.64 meV/A$^2$ for MCu7. Note that Figure 6 shows distinct differences in the mechanical response for the "low SFE" cases and "high SFE" cases identified above. While not easily seen, there are much smaller distinctions between the behaviors within those two groups.

In both the low and high SFE cases at the strain levels up to 5%, the active slip systems correspond to the ABC ($A^T B^T C^T$) and ADB ($A^T D^T B^T$) slip planes of the double Thompson tetrahedron (see Fig. 7). These slip planes are, respectively, parallel to the twin boundary planes and inclined to the twin boundary planes, and they intersect along a line that lies perpendicular to the tensile axis. The dislocations we observe acting on these slip planes are the extended 60° full dislocations [19]. A dislocation of this type splits into a 30° leading partial, an intrinsic stacking fault and a 60° trailing partial. The splitting distance between the partials, $r$, naturally depends on the SFE – it increases when the SFE decreases [1]. The splitting distance also depends on the resolved shear stress on the glide plane of the dislocation [1], so, under the high stress required to nucleate dislocations from the grain boundaries, the magnitude of $r$ can be significant and, for a small grain of size $d$, $r$ becomes compatible to, or even larger than $d$. This is important, because a complete extended dislocation cannot be nucleated unless $d > r$ [20].

Our main observations about the basic characteristics of the slip behavior for "low SFE" and "high SFE" potentials are summarized in Table II. In that table we summarize observations on the slip, stacking faults, ability of dislocations to penetrate the twin boundaries, the stability of the twin boundary structures and the formation of vacancies. The overall deformation behavior is a result of the interplay of all these phenomena. In the rest of this section, we comment in more detail on these observations.

*4.1. Slip by partials vs. complete dislocations*

Nucleation of isolated partial dislocations with trailing stacking faults is evidently preferred when the SFE is low. This observation is in agreement with the results obtained in [7] if we take into account that the unstable stacking fault energy is about the same for all semi-empirical potentials used in the present study.

The stress required to activate slip depends on the force applied to the dislocation, the lattice resistance force, and all other back-stresses. The force applied to the dislocation depends on the Peach-Koehler force, projected on the appropriate slip plane. The restraining forces include the lattice resistance ("Peierls-Nabarro") force, the curvature of the nucleating dislocation segment, and in the case of partial dislocations, the restraining effect of the trailing stacking fault. Since the forward stress on the dislocation depends on the Burgers vector, and the back-stress depends on the existence of a trailing stacking fault, complete lattice dislocations always experience larger forces under applied stress than partial dislocations do. The reason why partial dislocations nucleate, then, is because they have lower Peierls-Nabarro barriers to overcome, and we conclude that the magnitude of this barrier is related to the stacking fault energy.

*4.2. Slip transmission through grain or twin boundaries*

When dislocations glide on the ADB slip planes that are inclined to the twin boundary planes, they inevitably run into twin boundary planes. When they glide on the ABC slip planes, parallel to the twin boundaries, they eventually run into grain boundaries.

In the case of the potential MCu1, which has the smallest SFE, we did not observe any slip transmission through CTBs/GBs at strains up to ~5.5%. In the case of the potential MCu7, which has the highest SFE, the activation of the slip is delayed, compared to the case of the potential MCu1. However, since the splitting distance between the partials is much smaller in the MCu7 case, at higher strains we mostly observed complete extended dislocations. These dislocations crossed the grains, and interacted with or transmitted through the CTBs and GBs. In other words, in this case we observed slip transmission at much smaller strain levels, compared to the MCu1 case. The plastic deformation readily propagates from one grain to another, and, as a result, the peak stress (see Fig. 5) in the MCu7 case is significantly lower than that in the MCu1 case.

With the MCu1 potential, the first trailing partials nucleate at strain of ~6.0%, leading partials encounter twin boundaries and grain boundaries independently from their trailing partials, and extended dislocations are effectively blocked by the CTBs. This appears to correspond to the geometrical constraints on the transmission of partial dislocations that were described in [21]. At the strain of ~7.0% we observed the first events of slip transmission through the CTBs. At about the same strain, the Shockley partial dislocation loops were activated on the slip systems that correspond to ADC ($A^TD^TC^T$) and BDC ($B^TD^TC^T$) slip planes of the double Thompson tetrahedron (see Fig. 7). The sharp decrease of the stress on the stress-strain curve (see Fig. 5) corresponds to the strain level at which these Shockley partial dislocation loops are activated. We note that the results obtained using the other three EAM potentials with lower SFE (MCu2-MCu4) are qualitatively very similar to the results described throughout the text for the MCu1 case; the results obtained using the MCu6 potential (second largest SFE) are very similar to those obtained using the MCu7 potential (the largest SFE). The results obtained using the MCu5 potential indicate a transition between the two groups of "low SFE" cases (MCu1-MCu4) and "high SFE" cases (MCu6 and MCu7). In the case of potentials MCu2-MCu4, the splitting distance between the partials is larger and the slip activation takes place at smaller strain levels, but the slip transmission is delayed, compared to the cases of the potentials characterized by the larger values of the SFE (MCu6 and MCu7).

*4.3. Twin boundary mobility*

The transmission of dislocations through twin boundaries creates interfacial defects (sometimes called "disconnections" or, in this specific case, "twinning dislocations") that have both dislocation and step character and whose motion promotes the lateral migration of the twin boundary. Thus, the deformation of high-SFE materials results in dramatically increased mobility of the twin boundaries. Interfacial steps can be seen in the snapshots from the MCu7 case in Fig. 6. Because these topological defects also have dislocation character, they can move in response to the applied stresses. We see that the initial twin spacings are significantly changed during the deformation process for the high-SFE cases but that the spacings are essentially unchanged for the low-SFE simulations, presumably as a result of the lack of interfacial defect creation in these cases.

*4.4 Quadruple junction formation*

In some cases, the migration of the twin boundaries in the high-SFE simulations results in the formation of quadruple junctions where two twin boundaries from adjacent columns migrate so that they meet along a line at the grain boundary. For example, in the MCu7 case quadruple junctions are formed at the strain of ~8.0%. In the process of the formation of quadruple junctions, the 54.94º ATGBs were completely eliminated, and the 35.47º STGBs transformed into the CTBs (see Fig. 6). The elimination of the 54.94º ATGBs was associated with the migration of the originally present CTBs. This migration process involved the formation of steps on the CTBs that were formed as the result of the transmission of dislocations through the CTBs, as discussed above. Interfacial steps on CTBs were also emitted from the triple-junctions between the 35.47º STGBs, 54.94º ATGBs and the CTBs. The migration of all of these steps results in the migration of the CTBs (see Fig. 6). We also observed the formation of quadruple junctions in the MCu6 case, which has the second largest SFE in the series of the developed potentials, but no quadruple junctions were observed for the potentials with smaller SFE.

Because the two boundary segments at each quadruple junction have different misorientations, in general they also have different energies; it remains unclear whether the formation of these quadruple junctions in our simulations is a reflection of local thermodynamic equilibrium or other factors, and this is a matter of further investigation.

*4.5. Vacancy formation*

The grey dots that build up in the images shown in Fig. 6 correspond to vacancies. It can be seen that the vacancy concentration builds to significantly higher levels at high strains in the high-SFE case. There is also evidence, in some cases, of the formation of strings of vacancies lying along slip planes, parallel to the active Burgers vectors. These are consistent with a mechanism in which jogs are formed by the intersection of gliding dislocations on different slip planes which are then dragged by the moving dislocations, forcing them to move by climb and emit vacancies. The excess vacancy concentrations formed in the high-SFE cases may be expected to increase recovery rates in post-deformation annealing, or even allow a certain amount of recovery to occur at low temperatures.

*4.6. Length scale effects*

All of our simulations have been carried out using a scheme in which the twin widths (or twin boundary spacings) are initially the same. Marked differences appear between cases in which the operational stacking fault width (under applied stress) allows partials to interact independently with the CTBs ("low SFE" cases) and cases in which the leading and trailing partials interact collectively ("high SFE" cases).

In real nanotwinned materials, there is always a distribution of twin widths and if that distribution spans the operational stacking fault width, it is possible that the wider twins may operate much like the high SFE cases here, while the narrower ones operate in a way that is similar our low SFE cases. In effect, then, twin width and SFE may be seen as conjugate variables. Where we see sharp transitions and distinctions between high and low SFE cases, these transitions may appear considerably less sharp in real materials that have a distribution of twin thicknesses.

**5. Conclusions**

We have developed a set of potentials that allow us to study stacking fault energy as an isolated variable. This has been applied to the deformation of nanotwinned fcc materials, nominally copper, and we have identified several deformation behaviors that differ sharply between "high" and "low" SFE types (where the definition of "high" and "low" may depend on microstructural variables such as the distribution of twin widths).

In general, low SFEs result in:

  i. Independent reactions of Shockley partial dislocations;
 ii. Higher peak stresses associated with slip propagation across interfaces;
iii. Low CTB mobility;
 iv. Low vacancy production.

Conversely, high SFEs result in:

  i. Collective reactions of complete lattice dislocations;
 ii. Lower peak stresses associated with slip transmission across interfaces;
iii. Increased CTB mobility (allowing for the formation of quadruple junctions);
 iv. Accumulation of high vacancy concentrations.

The present study represents a testbed that can be adopted and/or extended to verify the validity of the underlying assumptions of analytical models which describe the effects of the stacking fault energy on the deformation mechanisms in nanotwinned face-centered cubic metals.

**Acknowledgements**

Work at the Ames Laboratory was supported by the Department of Energy, Office of Basic Energy Sciences, under Contract No. DE-AC02-07CH11358.

Table I. Properties calculated with EAM potentials developed in the present study[§].

| Property | MCu1 | MCu2 | MCu3 | MCu4 | MCu5 | MCu6 | MCu7 |
|---|---|---|---|---|---|---|---|
| $a$ (fcc) (Å) | 3.639 | 3.639 | 3.639 | 3.639 | 3.639 | 3.638 | 3.638 |
| $E_{coh}$ (fcc) (eV/atom) | -3.425 | -3.416 | -3.423 | -3.429 | -3.428 | -3.410 | -3.427 |
| $E_f^v$ (fcc) (eV/atom) | 1.11 | 1.12 | 1.11 | 1.10 | 1.07 | 1.05 | 1.03 |
| $E_f^m$ (fcc) (eV/atom) | 0.91 | 0.91 | 0.92 | 0.93 | 0.95 | 0.98 | 1.00 |
| $C_{11}$ (GPa) | 173 | 174 | 174 | 175 | 175 | 177 | 178 |
| $C_{12}$ (GPa) | 128 | 127 | 127 | 127 | 127 | 126 | 125 |
| $C_{44}$ (GPa) | 84 | 84 | 84 | 84 | 84 | 83 | 83 |
| $E_f^i$ (<100> fcc) (eV/atom) | 2.82 | 2.81 | 2.81 | 2.82 | 2.82 | 2.82 | 2.81 |
| $\gamma$(<100> fcc) (meV/Å$^2$) | 79 | 78 | 75 | 72 | 66 | 58 | 53 |
| $\gamma$(<110> fcc) (meV/Å$^2$) | 83 | 82 | 80 | 77 | 72 | 65 | 60 |
| $\gamma$(<111> fcc) (meV/Å$^2$) | 68 | 67 | 64 | 61 | 55 | 47 | 41 |
| $E^{SF}$ (meV/Å$^2$) | 0.91 | 1.55 | 2.76 | 3.85 | 5.91 | 9.35 | 11.65 |
| $E^{CTB}$ energy (meV/Å$^2$) | 0.46 | 0.78 | 1.38 | 1.92 | 2.96 | 4.68 | 5.82 |
| $\Delta E_{fcc \to bcc}$ (eV/atom) | 0.040 | 0.041 | 0.041 | 0.042 | 0.043 | 0.045 | 0.046 |
| $\Delta E_{fcc \to hcp}$ (eV/atom) | 0.0026 | 0.0044 | 0.0080 | 0.0109 | 0.0168 | 0.0266 | 0.0331 |
| $T_m$ (fcc, K) | 1349 | 1352 | 1353 | 1355 | 1356 | 1353 | 1351 |
| $\Delta H_m$ (fcc) (eV/atom) | 0.130 | 0.130 | 0.130 | 0.129 | 0.129 | 0.127 | 0.127 |
| $\Delta V_m$ (fcc) (%) | 3.5 | 3.4 | 3.4 | 3.4 | 3.5 | 3.5 | 3.5 |

[§]In this table, $a$ is the lattice parameter, $E_{coh}$ is the cohesive energy, $E_f^v$ and $E_f^m$ are vacancy formation and migration energies, respectively, $C_{ij}$ are the elastic constants, $E_f^i$ is the interstitial formation energy, $\gamma$, $E^{SF}$ and $E^{CTB}$ are the free surface, stacking fault and coherent twin boundary energies, respectively, $\Delta E_{fcc \to \beta}$ is the difference in energies of the β and fcc phases, $T_m$ is the melting temperature, $\Delta H_m$ and $\Delta V_m$ are the latent heat and change in the atomic volume upon melting.

Table II. Main observations from the MD simulations of plastic deformation in fcc nanotwinned materials.

| SFE | Low | High |
|---|---|---|
| Slip | Slip occurs by the nucleation and glide of Shockley partial dislocations. | Slip occurs by the motion of complete lattice dislocations. |
| Stacking faults | As strain builds up, more and more stacking faults span the twins and the full width of the grains modeled here. | Very few stacking faults are observed. |
| Ability of dislocations to penetrate TBs | Very few dislocations penetrate the twin boundaries. | Dislocations readily penetrate the twin boundaries, leaving steps on TBs that can move and allow the twin boundaries to migrate. |
| Stability of TB structure | TBs are immobile. | TBs can migrate because of steps left by the dislocation penetrating though them. |
| Formation of vacancies | Vacancy deposition by moving dislocations is relatively low. | Vacancy deposition by moving dislocations is significant. |

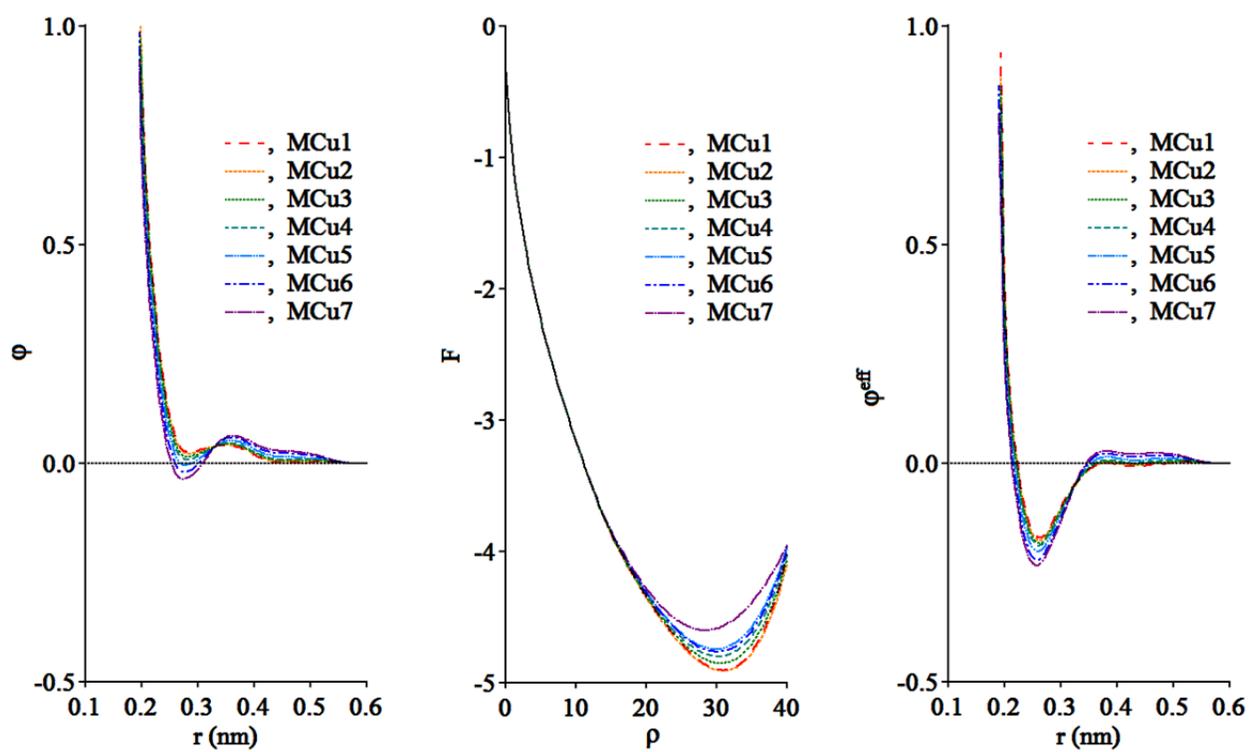

Figure 1. The potential functions (the density function is the same for all potentials).

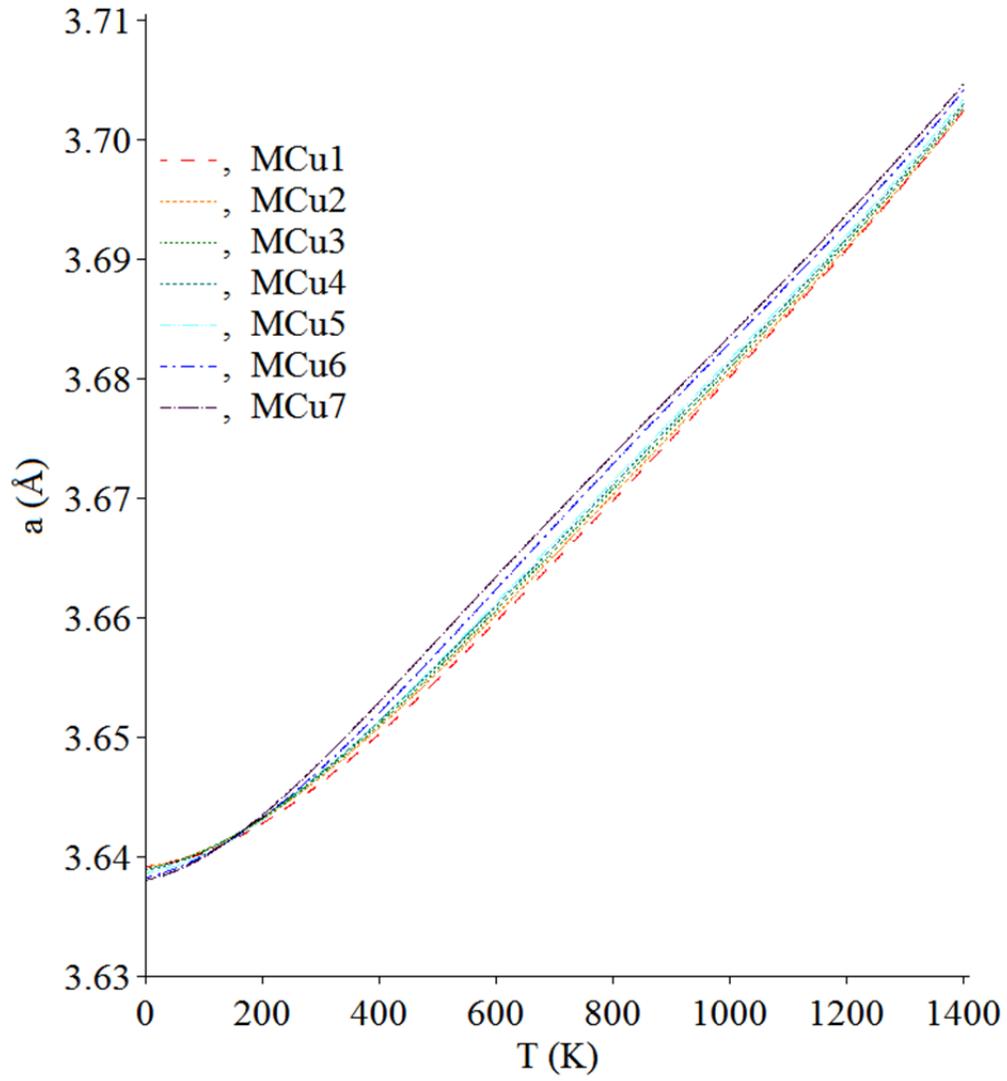

Figure 2. Lattice parameter as function of temperature.

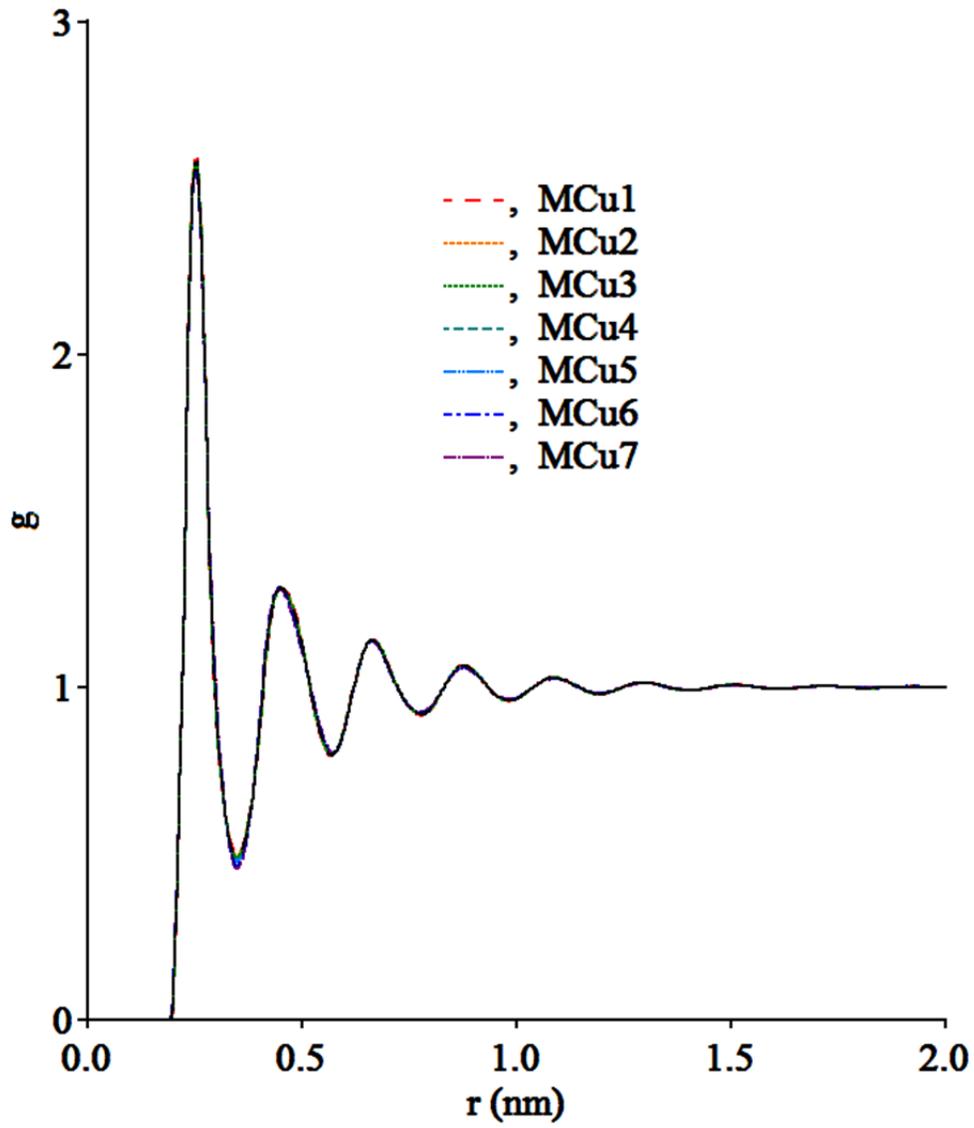

Figure 3. Liquid pair correlation functions at T=1973 K.

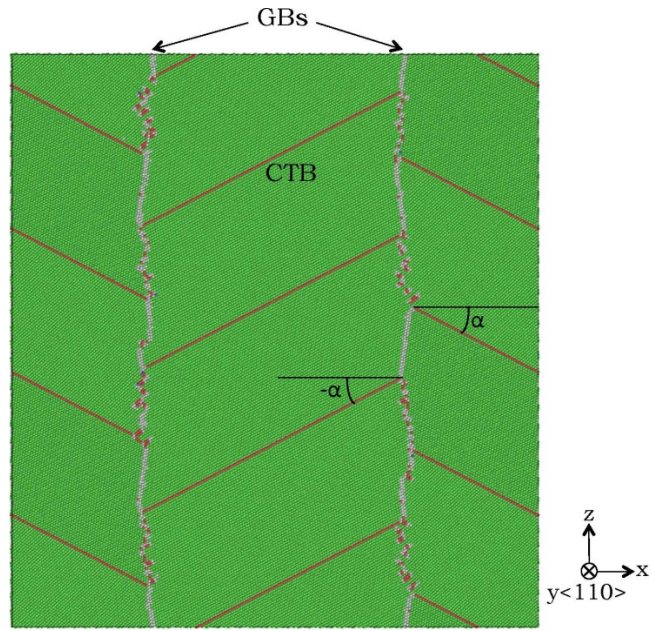

Figure 4. Simulation cell with atoms are colored according to CNA. The color-coding is as follows: green – FCC, red – HCP, grey – other.

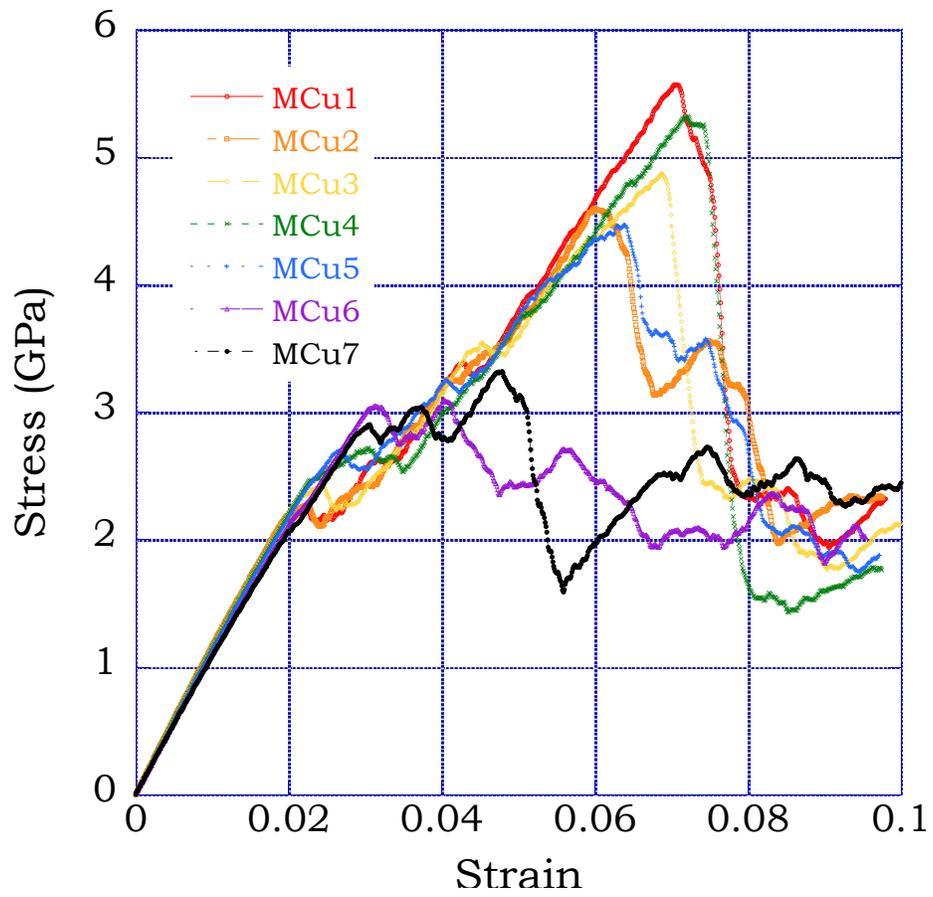

Figure 5. Stress-strain curves obtained using different potentials.

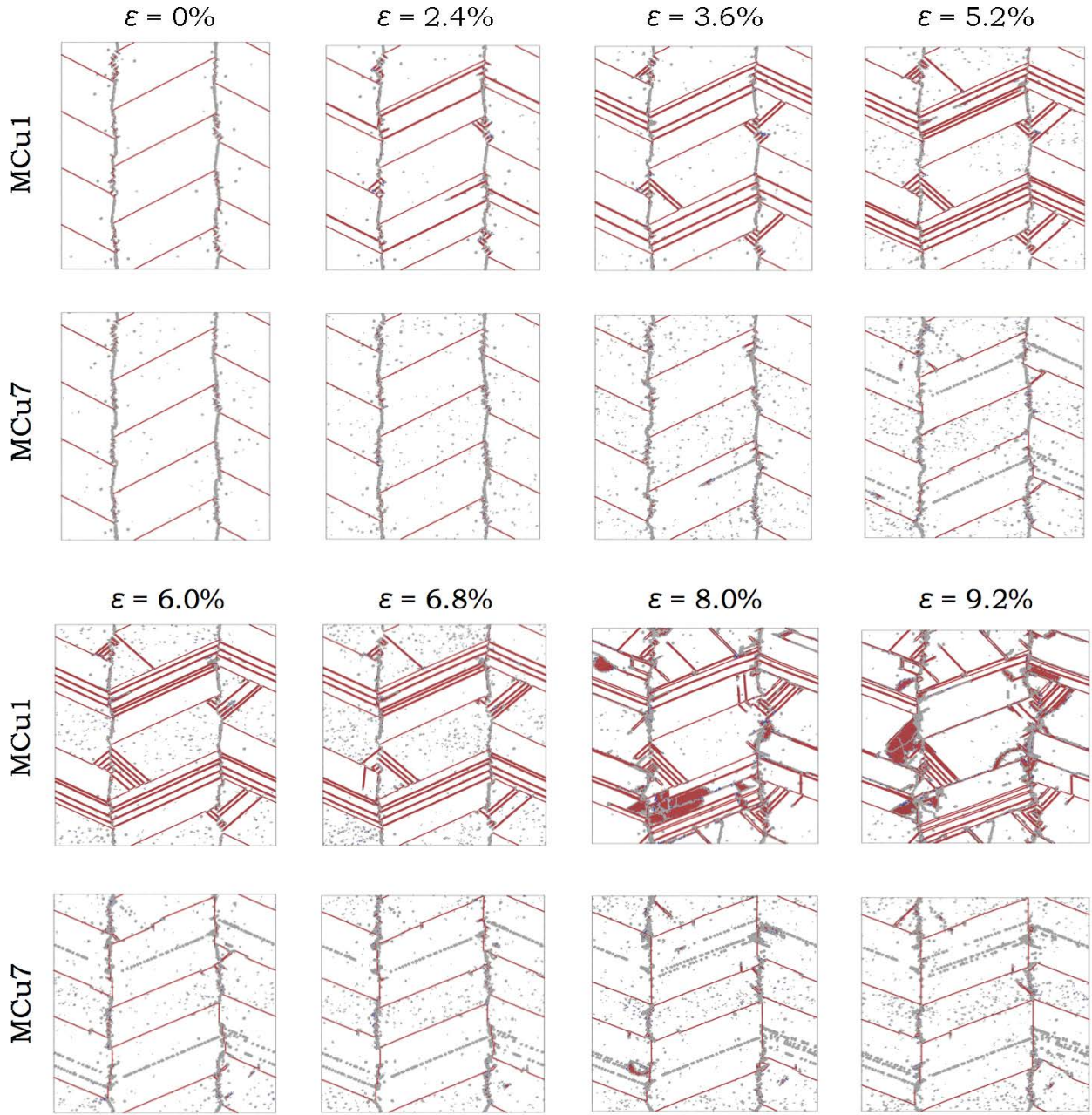

Figure 6. The snapshots corresponding to different strain levels, obtained using the EAM potentials MCu1 ($E^{SFE}$=0.91 meV/Å$^2$) and MCu7 ($E^{SFE}$=11.64 meV/Å$^2$). FCC atoms are not shown. The atoms are colored according to the CNA. The color-coding is as follows: red – HCP, blue – BCC, grey – other.

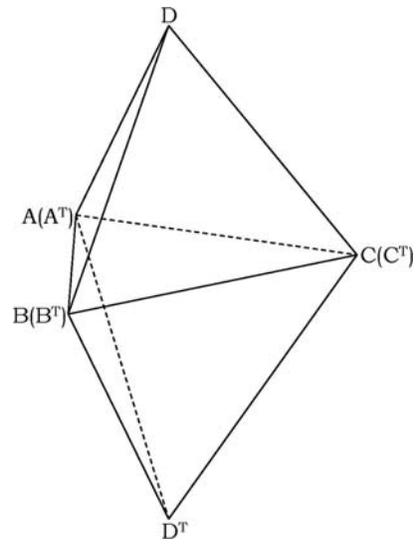

Figure 7. The fcc twin octahedron, formed by two Thompson tetrahedrons (in a matrix and twin systems).